# Characterization of Downlink Transmit Power Control during Soft Handover in WCDMA Systems

**Palash Gupta, Hussain Mohammed, and M.M.A Hashem**
Department of Computer Science and Engineering
Khulna University of Engineering and Technology (KUET)
Khulna-9203, Bangladesh
spline_palash@yahoo.com, redwan0207046@yahoo.com , hashem@cse.kuet.ac.bd

## Abstract

*This paper presents the characterization of power control in WCDMA systems. We know that CDMA is an interference limited system. It is shown that the unbalance scheme is reliable and successful for both 2-way and 3-way soft handover. Unbalance scheme minimizes interference and speed up the soft handover algorithm to support more users quickly. Furthermore it requires minimum time to make decision for proper power control in soft handover status.*

## I. INTRODUCTION

During soft handover, the power control procedure is more complicated because there are at least two BSs involved. Power control and soft handover are two essential techniques, which effectively increase the spectrum efficiency in CDMA systems. Power control aims at minimizing the total interference and soft handover handles the mobility of the mobile terminals. In the uplink direction, the mobile terminal adjusts it's transmit power based on the combination of received transmit power control (TPC) commands from all the base stations in the active set. Therefore, the reliability of TPC bits and the proper combining strategy are fatal to the uplink power control during soft handover [1], [2]. Differently, in the downlink direction, only one TPC command is sent by the mobile terminal. All the base stations in the active set adjust their transmit power based on this TPC command.

Several downlink power control during soft handover have been proposed [2], [3], [4], [5].The proposed scheme in this paper is an optimized scheme, base station transmits power during soft handover in the downlink direction of WCDMA FDD systems. The reason for focusing on the downlink is because the downlink is more likely to be the bottleneck for the third generation mobile systems due to the symmetric frequency band allocation but asymmetric load requirement between the downlink and the uplink. Several downlink power control schemes during soft handover have been proposed in literature. Two well known schemes for power control adopted by 3GPP is considered when necessary [6]. Both balance and unbalance scheme are not necessary to use in power control in soft handover status. In this paper we prove the superiority of the unbalance scheme and try to characterize the proposed optimized power control [6].

This paper is organized into six sections. Section 1 is the introduction. Section 2 describes the principles of the proposed power control scheme. In section 3 the feasibility of the changed scheme is evaluated. And in section 4 the system level performance is evaluated. Furthermore, in section 5 results are presented, followed by the conclusions in section 6.

## II. PPRINCIPLES OF OPTIMIZED DOWNLINK POWER CONTROL

We know that the base station changes their transmit power dynamically from both balance and unbalance scheme. Power is needed for mobile in soft handover status changes depending upon its location. To mobiles near the cell boundary balance scheme is better and to mobiles not near the cell boundary unbalance scheme is better [6]. We consider three-way soft handover.

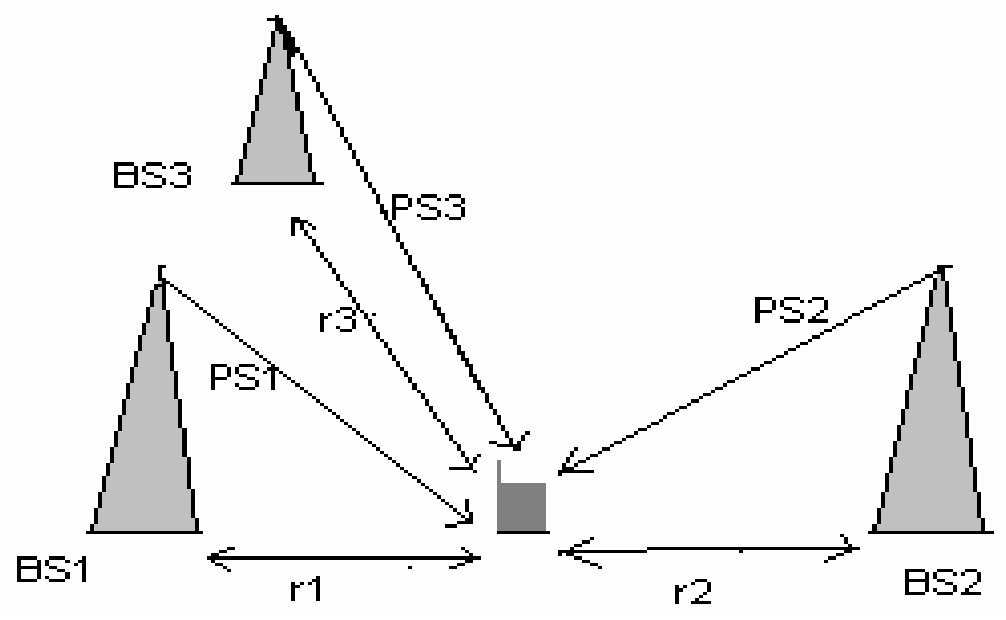

Fig. 1: Downlink power control during soft handover

Fig. 1 shows a mobile in soft handover status. Assuming a uniform load distribution within the system, the total transmitted power of all the base stations are the same, defined as $P_{total}$. $Ps_1$, $Ps_2$ and $Ps_3$ are the transmit power to the mobile from its serving base stations $BS_1$, $BS_2$ and $BS_3$ separately. After maximal ratio combining the received bit energy-to-interference power spectral density ratio $E_b/I_o$ of the mobile is as follows

$$\frac{E_b}{I_0} = \left[\frac{E_b}{I_0}\right]_1 + \left[\frac{E_b}{I_0}\right]_2 + \left[\frac{E_b}{I_0}\right]_3 \quad (1)$$

Where

$$\left[\frac{E_b}{I_0}\right]_1 = \frac{WP_{s1}L_1}{vR\left(P_{total}-P_{s1}\right)\left(1-a\right)L_1+\sum_{i\neq 1}P_{total}L_i}$$

$$= \frac{WP_{s1}}{vRP_{total}\left(1-a+\sum_{i\neq 1}\frac{L_i}{L_1}\right)-P_{s1}\left(1-a\right)} \quad (2)$$

Where $W$ is the chip rate; $R$ is the service bit rate; $v$ is the activity factor of the service; $a$ is the downlink orthogonality factor with 1 for perfect orthogonality and 0 for nonorthogonality; $L_i$ is the propagation attenuation from $BS_i$ to the mobile; summation of $\sum P_{total}L_i$ represents the inter-cell interference received by the mobile from all the base stations except for $BS_1$. $[E_b/I_0]_2$ and $[E_b/I_0]_3$ can be derived similarly. Usually mobiles in soft handover status are located near the cell boundary. To these mobiles, $Ps_1(1-a)$ is negligible compared to the total interference. Thus, (1) can be written as

$$\frac{E_b}{I_0} \approx \frac{W}{vRP_{total}}\left[\frac{P_{s1}}{1-a+\sum_{i\neq 1}\frac{L_i}{L_1}}+\frac{P_{s2}}{1-a+\sum_{j\neq 2}\frac{L_j}{L_2}}+\frac{P_{s3}}{1-a+\sum_{k\neq 3}\frac{L_k}{L_2}}\right] \quad (3)$$

In the perfect situation $Ps_1 = Ps_2 = Ps_3$ .Using the assumption, the transmit power for each downlink channel can be expressed as

$$Ps_1 = Ps_2 = Ps_3 = \frac{\left(\frac{E_b}{I_0}\right)_t \frac{vR}{W}P_{total}}{\frac{1}{1-a+\sum_{i\neq 1}\frac{L_i}{L_1}}+\frac{1}{1-a+\sum_{j\neq 2}\frac{L_j}{L_1}}+\frac{1}{1-a+\sum_{k\neq 3}\frac{L_k}{L_1}}} \quad (4)$$

Define $B$ and $P_t$ as

$$B = \frac{P_{s1}}{P_{s2}} = \frac{P_{s2}}{P_{s3}}, P_t = P_{s1} + P_{s2} + P_{s3} \quad (5)$$

In order to guarantee the quality of service (QoS), the total transmitted power to this certain mobile can be obtained by substituting (4) into (3), Shown as (5)

$$P_t = P_{s1}+P_{s2}+P_{s3} \approx \left(1+\frac{1}{B}+\frac{1}{B}\right)\frac{\left(\frac{E_b}{I_0}\right)_t \frac{vR}{W}P_{total}}{\frac{1}{1-a+\sum_{i\neq 1}\frac{L_i}{L_1}}+\frac{1}{B\left(1-a+\sum_{j\neq 2}\frac{L_j}{L_2}\right)}+\frac{1}{B\left(1-a+\sum_{k\neq 3}\frac{L_k}{L_2}\right)}} \quad (6)$$

Where $\left(E_b/I_0\right)_t$ is the target value of required $E_b/I_0$. CDMA systems are interference limited system. Minimizing the total interference is the basic principles for optimizing the radio resource. Here the optimized scheme fulfills the entire criterion. A proper ratio of B is selected from previous unbalance scheme [6]. So for unbalance scheme $B = L_1/L_2 = L_2/L_3$ , where $L_1$ , $L_2$ and $L_3$ is the propagation attenuation of the serving base stations. When the power ratio between the BSs in the active set equals the propagation attenuation ratio between the BSs, the total power consumption during the soft handover can be reduced compare to the balance power division scheme where $B = 1$ .

For the 2-way soft handover the power calculation is very similar to [7].

## III. FEASIBILITY EVALUATION

From (6), it is clear that the total transmit power $P_t$ is a function of power ratio $B$ and the propagation attenuation $L$. Using the standard propagation model in [4], $Li$ can be expressed as

$$L_i = r_i^{-\alpha} .10^{\zeta_i/10} \quad (7)$$

Where $\alpha$ is the path loss exponent and $\zeta_i$ (in dB) follows a Gaussian distribution, representing the attenuation due to shadowing from $i^{th}$ BS, with zero mean and a standard deviation of σ. Therefore, $L_i$ is related to the radio environment and the location of the mobile as well. Fig. 2 shows the average relative total transmits power $P_t/P_{total}$ to a mobile as a function of the different shadowing and mobile is not situated near the cell boundary means $r/R = 0.6$. And we consider $r/R = 1.0$ for mobiles near the cell boundary in Fig. 3. The parameter values are taken from practical ranges of about $\alpha$ =4, $v$=0.5, $a$=0.6, $\left(E_b/I_0\right)_t$=5 dB for 12.2 AMR speech service and $w$=3.84Mchip/s. $R$ is the radius of the cell, consider $R = 1$ . $r/R$ shows the relative distance from the mobile to one of the serving base stations. And 19 neighboring cells are considered for inter-cell interference. The typical value of σ lies between 8 to 10 [7]. We use 6-10 dB for simulation in Fig. 2, Fig. 3, Fig. 4 and Fig. 5 considering fluctuation in radio environment.

From the Fig. 2 and Fig. 4, we can say that for various shadowing loss unbalance scheme is better if mobile is not near from the cell boundary. From the Fig. 3 and Fig. 5 we can say that for various shadowing loss unbalance scheme is better till now if mobile is near from the cell boundary. $P_t/P_{total}$ is sensitive to the shadow fading. For that reason we check the relative transmit for different shadowing. From the two schemes, the better one is accepted to minimize the interference according to (5).

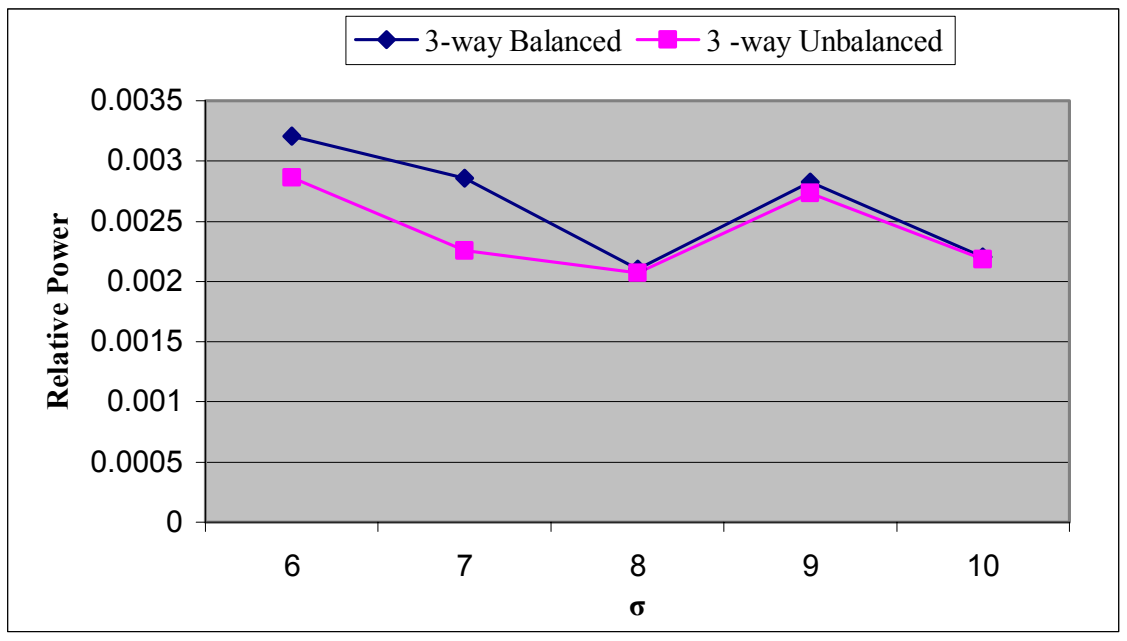

Fig. 2: Relative total transmit $P_t/P_{total}$ power for mobile depending on various shadowing (dB),where $r/R = 0.6$ in 3-way soft handover

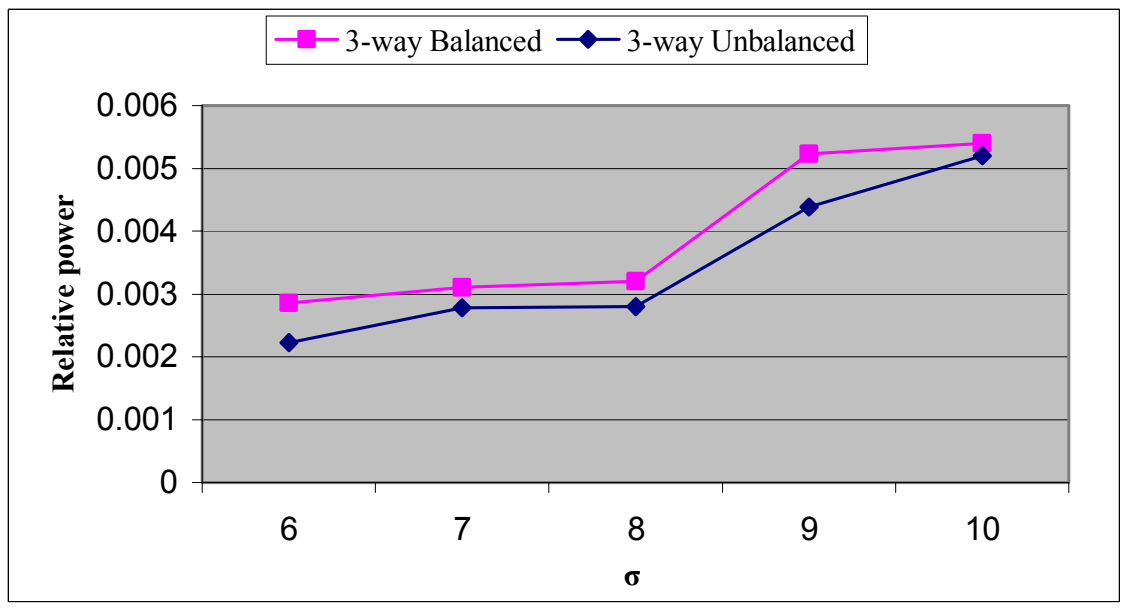

Fig. 3: Relative total transmit power $P_t/P_{total}$ for mobile depending on various shadowing (dB), where $r/R = 1.0$ in 3-way soft handover

From the Fig. 2 and Fig. 3 we can realize the superiority of the unbalance scheme for 3-way soft handover.

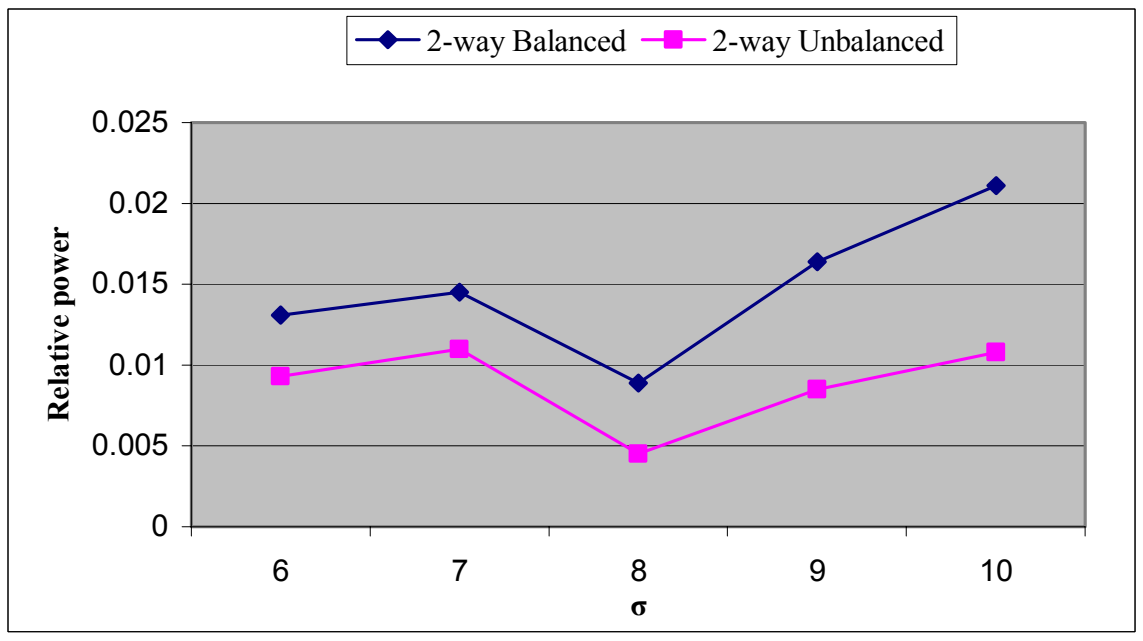

Fig. 4: Relative total transmit $P_t/P_{total}$ power for mobile depending on various shadowing (dB), where $r/R = 0.6$ in 2-way soft handover

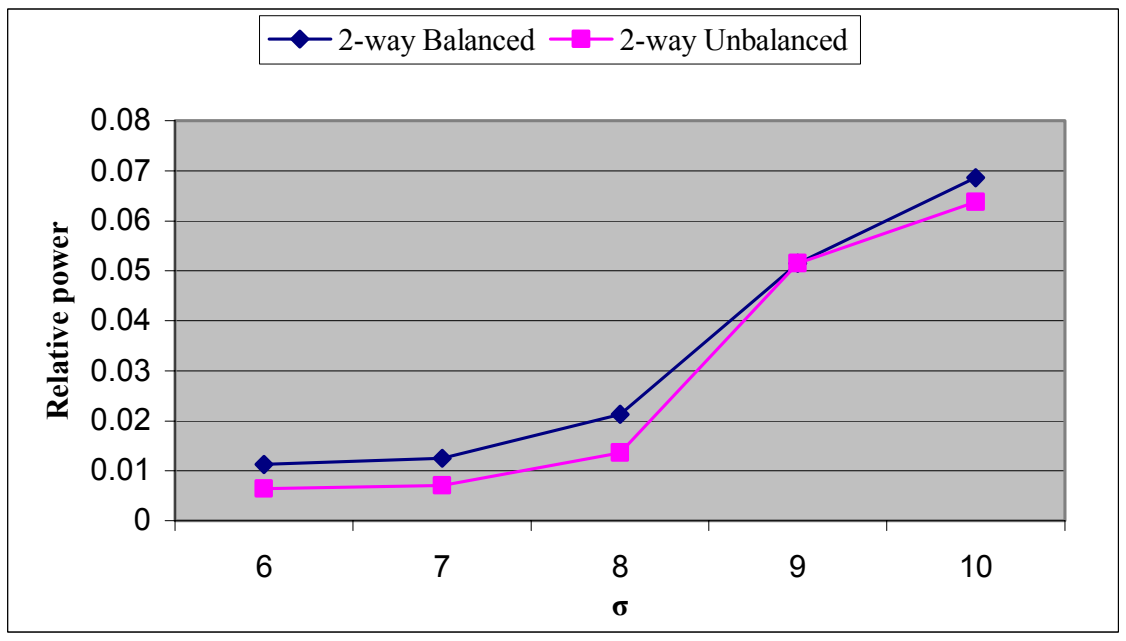

Fig. 5: Relative total transmit power $P_t/P_{total}$ for mobile depending on various shadowing (dB), where $r/R = 1.0$ in 2-way soft handover

So from the above Fig. 4 and Fig. 5 we can realize the superiority of the unbalance scheme for the 2-way soft handover. Using two schemes at a time is not actually efficient for bursty traffic [6]. It takes too much time for deciding power control scheme and actually its feasibility is unrealistic. This guarantees the feasibility of unbalance power allocation

## IV. SYSTEM LEVEL PERFORMANCE

In this section, the system level performance of the power control system is evaluated during soft handover. The downlink capacity gains caused by soft handover are compared with different power control schemes. We use the method proposed in previous work [8], [9] for analyzing the downlink capacity gain caused by soft handover. The system model is a WCDMA system with ideal hexagonal cell structure, uniform distribution of users and single type of service supported. We divide the actual coverage of the base station into three parts as shown in Fig. 6.

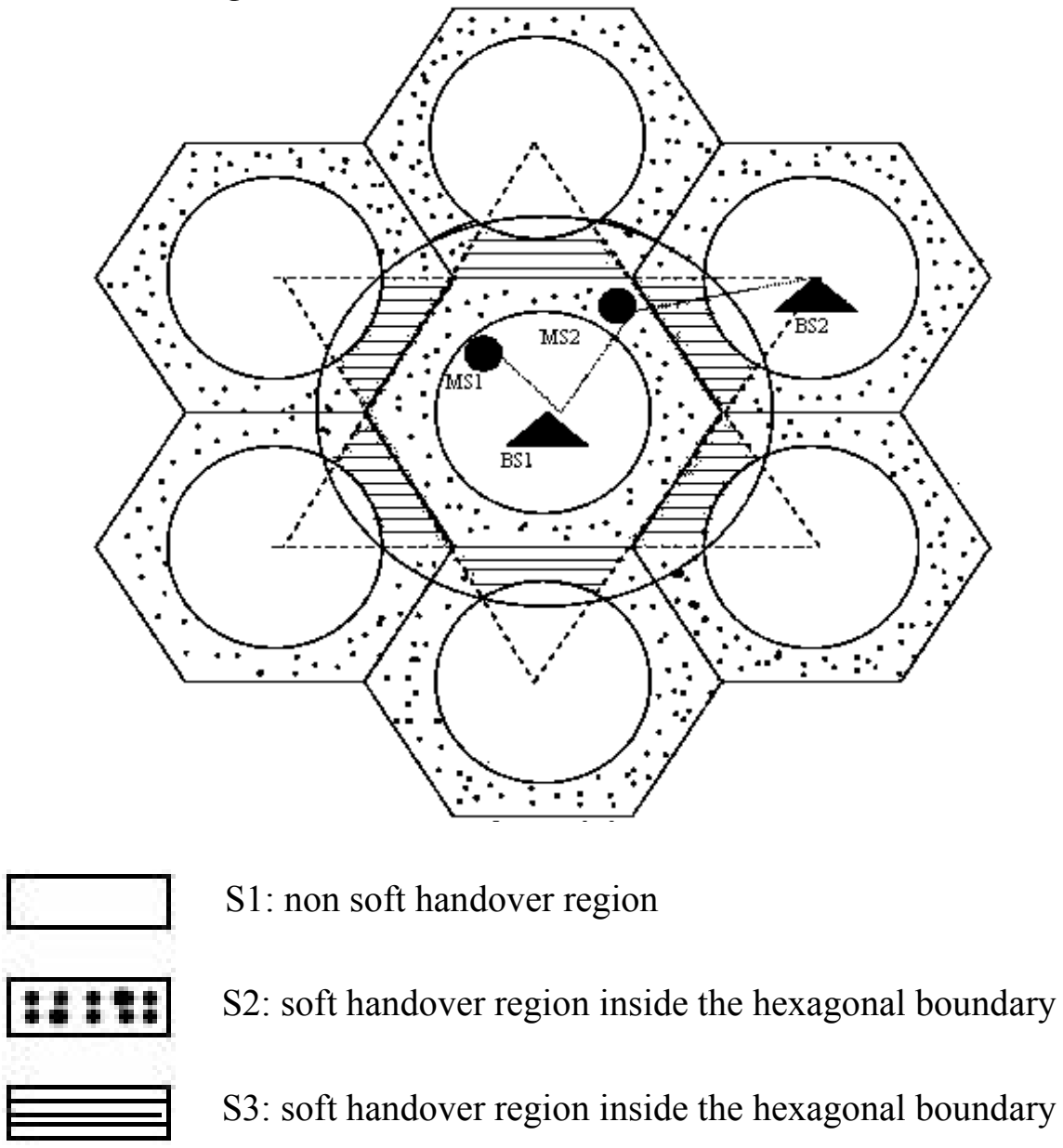

Fig. 6: Base station coverage

According to Fig. 6 consider, $S_1$ means the non soft handover region, $S_2$ or $S_3$ means the inside the soft handover region inside the hexagonal boundary. Furthermore, we consider a mobile station denoted by $MS_1$ is in outside the soft handover region and another mobile station denoted by $MS_2$ is in inside the soft handover region.

In this paper, we use UTRA soft handover algorithm [6]. We assume that all the base stations allocate the same amount of transmit power to the common pilot channel. To a mobile outside the soft handover region, as $MS_1$ in $S_1$, the required transmit power for the downlink dedicated channel $P_{1_out}$ can be expressed as:

$$P_{1_out} = \left(\frac{E_b}{I_0}\right)_t \frac{vR}{w} P_{total}\left[1 - a + \sum_{i=2}^{M}\frac{L_i}{L_1}\right] \tag{8}$$

To a mobile inside the soft handover region, as $MS_2$ in $S_2$ or $S_3$, the power level of $P_{1_in}$ is related to the power control schemes used during the soft handover:

For Balanced scheme

$$P_{1_in} = \frac{\left(\frac{E_b}{I_0}\right)_t \frac{vR}{W} P_T}{\frac{1}{1-a+\sum_{i=2}^{M}\frac{L_i}{L_1}} + \frac{1}{1-a+\sum_{\substack{j=1\\(j\neq k)}}^{M}\frac{L_j}{L_k}}} \tag{9}$$

For Unbalanced scheme

$$P_{1_in} = \frac{\left(\frac{E_b}{I_0}\right)_t \frac{vR}{W} P_{total}}{\frac{1}{1-a+\sum_{i=2}^{M}\frac{L_i}{L_1}} + \frac{1}{\left(\frac{L_1}{L_k}\right)\left(1-a+\sum_{\substack{j=1\\(j\neq k)}}^{M}\frac{L_j}{L_k}\right)}} \tag{10}$$

The total transmit power for $BS_1$ can be expressed as:

$$P_T = (1-\gamma)P_T + \iint_{s1} \rho P_{1_out} ds + \iint_{s1+s2} \rho P_{1_in} ds \quad \text{where}$$

$$\rho = \frac{2N}{3\sqrt{3}R^2} \tag{11}$$

Where $\gamma$ is the fraction of the total transmit power devote to the dedicated channels; $\rho$ is the density of users; $N$ is the number of active users per cell, $R$ is the radius of the cell.

## V. RESULT AND DISCUSSION

Substituting (8) , (9) and (10) to (11), the downlink capacity can be obtained. The results are based on the UTRA soft handover algorithm and normal cell selection with threshold CS_th equals 5db.

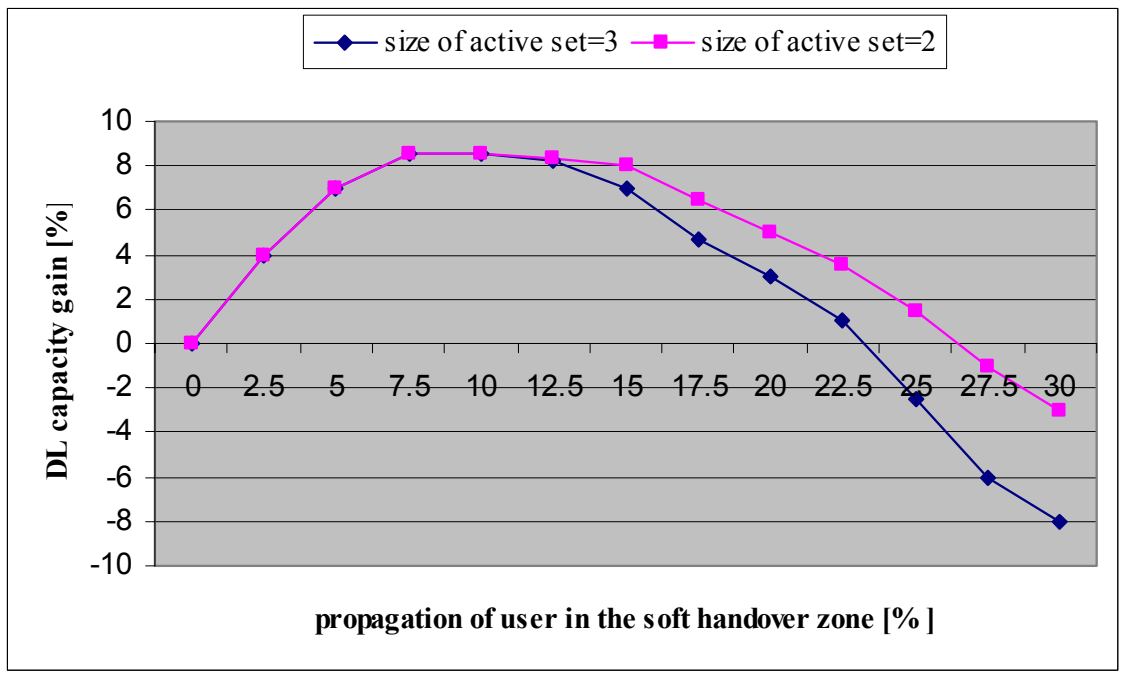

Fig. 7: soft handover gain with different size of active set

In Fig. 7, when the proportion of users in soft handover status is small ,there is no much difference between the two cases A(active set=2) and B(active set=3).However the proportion of the user in soft handover increases ,the performance of case B is worse than case A because there is too much interference being added. Therefore, considering the complexity and the increased signaling that comes with adding an extra BS in the active set, when implementing soft handover, the size of the active set should be kept two.

In Fig.8, the average downlink capacity gain as a function of soft handover overhead with different power control scheme is shown. The capacity gain is obtained by Two-way UTRA soft handover and normal cell selection with CS_th equals 5db.

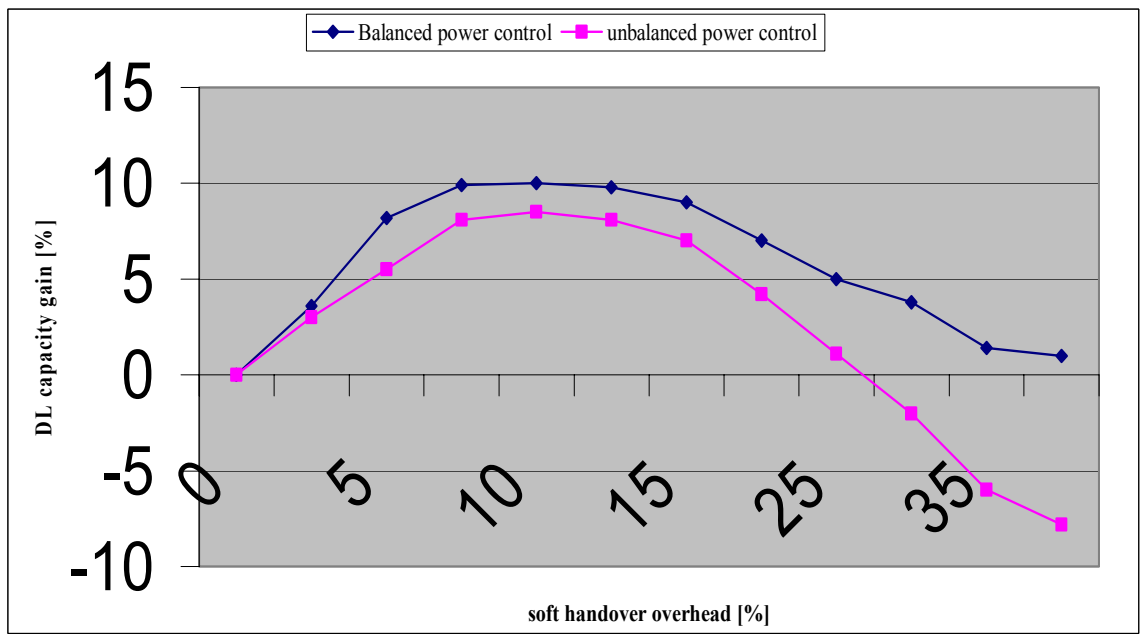

Fig. 8: Downlink capacity gain with different power control schemes

It is clear that when the soft handover is fixed, the downlink capacity gain is higher with the unbalance scheme. From the above feasibility evaluation section and in case of downlink capacity gain, the unbalance scheme minimizes the power and maximizes the downlink capacity. If we consider the algorithmic complexity, we need not take decision that which scheme gives relatively low power during soft handover. So using one scheme, extra comparison between two schemes is not necessary for power control. It is a very important point for bursty traffic to satisfy more users demand quickly.

## VI. CONCLUSIONS

We have calculated relative transmit power for various shadowing loss and find that both for near and far the unbalance scheme is better than balance. The result is same for both 2-way and 3-way soft handover. As the size of the active set should be kept to two because adding a BS to the active set raises interference, we can use unbalance scheme as a power control procedure in 2-way soft handover. Furthermore, we simulate the downlink capacity for both scheme and able to minimize the complexity of soft handover algorithm by using only the unbalance scheme, but not using the both.

Future work for the soft handover can be carried out by investigating the soft handover effects on bursty traffic.

## REFERENCES

[1] S. Park, D. K. Keun, C. S. Kang and S. M. Shin, “Uplink Transmit Power Control during Soft Handoff in DS?CDMA Systems,” Proceedings of VTC’2001, IEEE VTS 53rd, vol. 4, pp. 2913-2917, Spring 2001.

[2] B. Hashem and E. L. Strat, “On the Balancing of the Base Station Transmitted Powers during Soft Handoff in Cellular CDMA Systems,” Proceedings of ICC’2000, vol. 3, pp. 1497-1501, 2000.

[3] K. Hamabe, “Adjustment loop Transmit Power Con trol during Soft Handover in CDMA Cellular Systems,” Proceedings of VTC’2000, IEEE VTS 52$^{nd}$, vol. 4, pp. 1519-1523, Fall 2000.

[4] A. A. Daraiseh and M. Landolsi, “Optimized CDMA Forward Link Power Allocation During Soft Handoff,” Proceedings of VTC’98, IEEE VTS 48$^{th}$, vol. 2, pp. 1548-1552, 1998.

[5] H. Furukawa, K. Hamage and A. Ushirokawa, ”SSDT-site selection diversity transmission power control for CDMA forward link, ”IEEE J. Select. Areas Commun, vol. 18, No. 8, pp. 1546-1554, Aug 2000.

[6] Y. Chen and L. G. Cuthbert, “Downlink Perform ance of Different Soft Handoer Schemes for UMTS Systems” Proceedings of the international conference on Telecommunications ICT’2002, June, China, 2002. IEEE J. Select. Areas Commun., Vol. 4, No. 8, pp. 1281-1288, 1994.

[7] Y. Chen and L.G. Cuthbert, “Optimized Downlink Transmit Power Control during Soft Handover in WCDMA Systems”, *IEEE Wireless Communications and Networking Conference. WCNC2003*, March 2003.

[8] Y. Chen and L. G. Cuthbert, “Downlink Soft Handover Gain in UMTS Networks Under Different Power Control Conditions, ” Proceeding of the third international conference on Mobile Communication Technologies, 3G2002, pp. 47-51, London, UK, March, 2002

[9] A. Viterbi, A. M. Viterbi, K. S. Gilhousen, and E. Zehavi, “Soft handoffs extends CDMA coverage and increase reverse link capacity,” IEEE J. Select. Areas Commun., Vol. 4, No. 8, pp. 1281-1288, 1994.

.